\documentclass[preprint]{aastex}
\usepackage{subeqnarray}
\begin{document}
% Special additions for the PhD thesis.
\setlength{\textheight}{45\baselineskip}
% End of special additions.
\title{ 
Master Equation for Hydrogen Recombination
on Grain Surfaces
}
\author{
{Ofer Biham and Itay Furman \\ 
Racah Institute of Physics, The Hebrew University, Jerusalem 91904, 
Israel\\
Valerio Pirronello \\
Dipartimento di Metodologie Fisiche e Chimiche per l'Ingegneria, \\
Universita' di Catania, 95125 Catania, Sicily, Italy \\
and \\
Gianfranco Vidali \\
Department of Physics, Syracuse University, Syracuse, NY 13244}
}

\begin{abstract}
Recent experimental results on the formation of molecular hydrogen on
astrophysically relevant surfaces under conditions similar to those
encountered in the interstellar medium provided useful quantitative
information about these processes. Rate equation analysis of experiments
on olivine and amorphous carbon surfaces provided the activation energy
barriers for the diffusion and desorption processes relevant to hydrogen
recombination on these surfaces. However, the suitability of rate
equations for the simulation of hydrogen recombination on interstellar
grains, where there might be very few atoms on a grain at any given time,
has been questioned. To resolve this problem, we introduce a master 
equation that takes into account both the discrete nature of the H atoms
and the fluctuations in the number of atoms on a grain. The hydrogen
recombination rate on microscopic grains, as a function of grain size and
temperature, is then calculated using the master equation. The results are
compared to those obtained from the rate equations and the conditions
under which the master equation is required are identified.
\end{abstract}


\newpage

\section{Introduction}
        \label{intro.}

The formation of molecular hydrogen in the interstellar medium (ISM)
is a process of fundamental importance
\citep{Duley1984,Williams1998}.  
It was recognized long ago \citep{Gould1963} 
that ${\rm H}_2$ cannot form in the gas phase 
efficiently enough to account for its abundance.
It was thus proposed that dust grains act as catalysts,
where an H atom approaching the surface of a grain has
a probability $\xi$ to become adsorbed.
The adsorbed H atom (adatom)
spends an average time $t_{\rm H}$ (residence time)
before leaving the surface.
If during the residence time the H adatom encounters another H adatom, 
an ${\rm H}_2$ molecule will form with a certain probability.
Various aspects of this process were addressed in extensive theoretical 
studies
\citep{Gould1963,Williams1968,Hollenbach1970,Hollenbach1971a,Hollenbach1971b,Smoluchowski1981,Smoluchowski1983,Aronowitz1985,Duley1986,Pirronello1988,Sandford1993,Takahashi1999,Farebrother1999}.
In particular, Hollenbach et al.
calculated the sticking and mobility of
H atoms on grain surfaces.
They concluded that tunneling between adsorption sites, 
even at 
temperature as low as
$T=10$K, provides the required mobility.
The steady state production rate of molecular hydrogen per unit volume
was expressed according to \citep{Hollenbach1971b}
\begin{equation}
\label{eq:salpeter}
	{\cal R}_{\rm H_2} = {1 \over 2}
		\rho_{\rm H} v_{\rm H} \sigma \gamma \rho_{\rm g},
\end{equation}
where 
$\rho_{\rm H}$ 
and 
$v_{\rm H}$ 
are the number density and the speed
of H atoms in the gas phase, respectively,
$\sigma$ is the average cross-sectional area of a grain and
$\rho_{\rm g}$ 
is the number density of dust grains.
The parameter
$\gamma$ is the fraction of H atoms striking the grain
that eventually form a molecule, namely $\gamma = \xi \eta$,
where
$\eta$ 
is the probability that an H adatom on the surface
will recombine with another H atom
to form 
${\rm H}_2$.

Recently, 
a series of experiments were conducted 
to measure
hydrogen recombination
in an ultra high-vacuum (UHV) chamber by
irradiating the sample 
with two beams of H and D atoms  
and monitoring the HD production rate
\citep{Pirronello1997a,Pirronello1997b,Pirronello1999}.
Two different substrates have been used: 
a natural olivine (a polycrystalline silicate containing
Mg$_2$SiO$_4$ and Fe$_2$SiO$_4$) slab 
and an amorphous carbon sample.
The substrate temperatures 
during hydrogen irradiation
were in the range 
between 5 K and 15 K.
The HD formation rate was measured using 
a quadrupole mass spectrometer
both 
during 
irradiation 
and in a subsequent
temperature programmed desorption
(TPD) experiment in which the sample temperature 
was quickly ramped to over 30 K
to desorb all weakly adsorbed species.
It was found that H and D
atoms adsorbed on the surface
at the lowest irradiation temperature of 5 K
form molecules
during TPD
only above 9 K in the case of olivine 
and above 14 K
in the case of amorphous carbon.
This indicates that
tunneling 
alone does not provide enough mobility to H adatoms to enable 
recombination, and thermal activation is required.
The experimental results 
were analyzed 
using a rate equation model
(Katz et al. 1999).
In this analysis
the parameters of the rate equations
were fitted to the experimental TPD curves.
These parameters are
the activation energy barriers for atomic
hydrogen diffusion and desorption, 
the barrier for molecular hydrogen
desorption and the probability of spontaneous desorption of a hydrogen
molecule upon recombination.  
Using the values of the
parameters that fit best the experimental results, 
the efficiency of
hydrogen recombination on the olivine and amorphous carbon surfaces 
was calculated
for interstellar conditions using
the same rate equation model.
It was found that for both samples the recombination efficiency is
strongly dependent on temperature and exhibits a narrow window of
high recombination efficiency along the temperature axis.

It was recently pointed out that since hydrogen recombination in the
interstellar space takes place on small grains, rate
equations have a limited range of validity
\citep{Tielens1995,Charnley1997,Caselli1998,Shalabiea1998}. 
This is due to the fact that these equations take into
account only average concentrations and ignore fluctuations as
well as the discrete nature of the H atoms.
These properties become significant in the limit of very small grains
and low incoming flux of H atoms, exactly the conditions encountered in 
diffuse interstellar clouds where hydrogen recombination on silicate and
carbon surfaces is expected to be relevant.
As the number of H atoms on a grain fluctuates in the range of 0, 1 or 2,
the recombination rate cannot be obtained from the average 
number alone.
This can be easily understood, since the recombination process requires at
least two H atoms simultaneously on the surface.
Comparisons with Monte Carlo simulations have shown that the rate
equations tend to over-estimate the recombination rate.
A modified set of rate equations 
which exhibits better agreement with Monte Carlo simulations
was introduced by Caselli et al.
(1998) and applied by Shalabiea et al. (1998) to a variety of chemical
reactions.
In these equations the rate coefficients are modified in a semi-empirical
way to take into account the effect of the finite grain size on the
recombination process.

In this paper we introduce a master equation that 
is particularly suitable for the simulation of chemical reactions on
microscopic grains.
It takes into account both the discrete nature of the
H atoms as well as the fluctuations. 
Its dynamical variables are the
probabilities $P_{\rm H}(N_{\rm H})$ 
that there are 
$N_{\rm H}$ 
atoms
on the grain
at time $t$. 
The time derivatives  
$\dot{P}_{\rm H}(N_{\rm H})$, $N_{\rm H}=0, 1, 2, \dots$
are expressed 
in terms of the adsorption, reaction and desorption terms.
The master equation provides the time evolution of 
$P_{\rm H}(N_{\rm H})$, 
$N_{\rm H}=0, 1, 2, \dots$, 
from which
the recombination rate can be calculated.
We use it
in conjunction with the surface parameters obtained experimentally,
to explore the hydrogen recombination process on microscopic grains
for grain sizes, flux and surface temperatures 
pertinent to the conditions in the interstellar medium.

The paper is organized as follows. 
The rate equation model is 
described in Sec. 
\ref{sec:RateModel}.
The master equation is introduced in Sec.
\ref{sec:MasterModel}.
Computer simulations
and results 
for hydrogen recombination on microscopic grains
under interstellar conditions
are
presented in Sec. 
\ref{sec:Simulations}.
The case of more complex reactions  
involving
multiple species
is considered in Sec.
\ref{sec:Complex} 
and a summary in Sec. 
\ref{sec:Summary}.
 
\section{Rate Equations for H$_2$ Formation on Macroscopic Surfaces}
\label{sec:RateModel}

Consider an experiment in which a flux of H atoms 
is irradiated on the surface. 
If the temperature is not too low
H atoms that stick to the
surface 
perform hops as random walkers 
and recombine when they encounter one another.
Let $n_{\rm H}(t)$ (in monolayers [ML]) be the coverage 
of H atoms on the surface and $n_{\rm H_2}(t)$ (also in ML) 
the coverage of H$_{2}$ molecules
at time $t$. 
We obtain the following set of rate equations:
\begin{subeqnarray}
\label{eq:N}
{ {d{n_{\rm H}}} \over {dt}} & = & f_{\rm H} \cdot (1 - n_{\rm H} -  
n_{\rm H_2}) - W_{\rm H} n_{\rm H} - 2 a_{\rm H} n_{\rm H}^{2} 
\slabel{eq:N1} \\
{ {d{n_{\rm H_2}}} \over {dt}}  & = & 
\mu a_{\rm H} n_{H}^{2} - W_{\rm H_2} n_{\rm H_2}. 
\slabel{eq:N2}
\end{subeqnarray}
The first term on the right hand side of 
Eq.~(\ref{eq:N1}) 
represents the 
flux of H atoms
multiplied by
the Langmuir-Hinshelwood rejection term. 
In this scheme H atoms deposited on top of H atoms
or H$_{2}$ molecules already on the surface are rejected. 
The parameter
$f_{\rm H}$ represents the 
{\em effective} flux
of atoms 
(in units of $ML s^{-1}$), 
namely,
the 
(temperature dependent)
sticking coefficient
$\xi(T)$
of the bare surface
is absorbed into 
$f_{\rm H}$.
The second term in 
Eq.~(\ref{eq:N1}) 
represents the desorption of H atoms from the
surface. 
The desorption coefficient is 
\begin{equation}
W_{\rm H} =  \nu \cdot \exp (- E_{1} / k_{B} T)  
\label{eq:P1}
\end{equation}
where $\nu$ is the attempt rate 
(standardly taken to be $10^{12}$ s$^{-1}$), 
$E_{1}$ 
is the activation energy barrier for desorption 
of an H atom and $T$ is the temperature.
The third term in 
Eq.~(\ref{eq:N1}) 
accounts for the depletion of the H population
on the surface due to recombination into H$_{2}$ molecules, 
where
\begin{equation} 
a_{\rm H} =  \nu \cdot \exp (- E_{0} / k_{B} T) 
\label{eq:Alpha}
\end{equation}
is the hopping rate of H atoms on the surface
and $E_{0}$ is the activation energy barrier for H diffusion. 
Here we assume that diffusion occurs only by thermal hopping,
in agreement with recent experimental results
\citep{Katz1999}.
We also assume that there is no barrier for recombination. 
The first term on the right hand side of 
Eq.~(\ref{eq:N2}) 
represents the creation of H$_{2}$ molecules. 
The factor $2$ in the third term of 
Eq.~(\ref{eq:N1}) 
does not appear here since it
takes two H atoms to form one molecule. 
The parameter
$\mu$ represents the fraction of H$_{2}$ molecules
that remains on the surface upon formation, 
while a fraction of $(1-\mu)$ is spontaneously desorbed due 
to the excess energy released in the recombination process.
The second term in 
Eq.~(\ref{eq:N2}) 
describes the desorption of H$_{2}$ molecules. 
The desorption coefficient is 
\begin{equation}
W_{\rm H_2}  =  \nu \cdot \exp (- E_{2} / k_{B} T), 
\label{eq:P2}
\end{equation}
where $E_{2}$ is the activation energy
barrier for H$_{2}$ desorption.
The H$_{2}$ production rate $r_{\rm H_2}$
(ML s$^{-1}$) 
is given by:  
\begin{equation}
r_{\rm H_2}  =  (1-\mu) \cdot a_{\rm H} n_{\rm H}^{2} +
 W_{\rm H_2} n_{\rm H_2}. 
\label{eq:Production}
\end{equation}
This model can be considered as a generalization of 
the Polanyi-Wigner equation
[see e.g. Chan et al. (1978)].
It provides a 
description of both first order 
and second order 
desorption kinetics 
for different regimes of temperature and flux. 

The model described by
Eqs.~(\ref{eq:N})
was used 
by Katz et al. (1999)
to analyze the results
of the TPD experiments 
\citep{Pirronello1997a,Pirronello1997b,Pirronello1999}.
The values of the parameters 
$E_0$, 
$E_1$, 
$E_2$, 
and
$\mu$,
that best fit the experimental results were obtained.
For the olivine sample it was found that
$E_0=24.7$ meV,
$E_1=32.1$ meV,
$E_2=27.1$ meV
and
$\mu=0.33$, 
while
for the amorphous carbon sample
$E_0=44.0$ meV,
$E_1=56.7$ meV,
$E_2=46.7$ meV
and
$\mu=0.413$.

The model 
[Eqs. (\ref{eq:N})]
was then used in order to
describe the steady state conditions 
that are reached when both the 
flux and the temperature are fixed. 
The steady state solution is then easily 
obtained by setting 
${d{n_{\rm H}} / dt} = 0$ 
and 
${d{n_{\rm H_2}} / dt} = 0$ 
and solving the 
quadratic equation for $n_{\rm H}$
\citep{Biham1998,Katz1999}.
In case that 
the Langmuir-Hinshelwood rejection term
can be neglected,
the steady-state coverages
are
\begin{subeqnarray}
\label{eq:steadysimp}
n_{\rm H}  &=&  
{ { - W_{\rm H} + 
\sqrt{ W_{\rm H}^{2} + 8 a_{\rm H} f_{\rm H}}  } \over {4 a_{\rm H}}}
\slabel{eq:steady1simp}	\\
n_{\rm H_2}  &=&  {\mu \over {8 a_{\rm H} W_{\rm H_2}} } 
\left({W_{\rm H}^2 + 4 a_{\rm H} f_{\rm H} - W_{\rm H}
 \sqrt{W_{\rm H}^2 + 8 a_{\rm H} f_{\rm H}}}\right). 
\slabel{eq:steady2simp}
\end{subeqnarray}
More complicated expressions are obtained when the rejection term
is taken into account
\citep{Katz1999}.
The recombination efficiency 
$\eta$
is defined as the 
fraction of the adsorbed H atoms that desorb in the form of H$_2$
molecules, namely
\begin{equation}
\eta = {r_{\rm H_2} \over {f_{\rm H}/2}}.
\end{equation}
Note that under steady state conditions $\eta$
is limited to the range 
$0 \le \eta \le 1$.

By varying the temperature and flux over the astrophysically relevant range 
the domain in which 
there is non-negligible recombination efficiency
was identified. 
It  was found that the
recombination efficiency is highly temperature dependent. 
For each of the two samples 
there is a narrow window of high efficiency along the temperature axis,
which shifts to higher temperatures as the flux is increased. 
For the astrophysically relevant flux range the
high efficiency temperature range for olivine 
was found to be between
$7 - 9$K,  
while for amorphous carbon it is between
$12 - 16$ K. 

Note that in steady state the dependence of the production rate $r_{\rm H_2}$
on $\mu$ and $W_{\rm H_2}$
is only through the Langmuir-Hinshelwood rejection term.
This is easy to understand since the parameter
$\mu$ only determines what fraction of the  H$_2$
will desorb upon formation and  what fraction will
desorb thermally at a later time.
The desorption rate $W_{\rm H_2}$
determines the coverage of H$_2$
molecules at steady state.
As long as the coverage of H and H$_2$ is low, 
the Langmuir-Hinshelwood rejection term is small and
$\mu$ and $W_{\rm H_2}$
have little effect on the production 
rate $r_{\rm H_2}$.
Under interstellar conditions the coverage is expected to be low.
Therefore, the master equation presented below, 
that is required only
when the number of atoms on the grain is small, 
does not
include the rejection term.
Note, however, that at lower temperatures in which H atoms are immobile
(thus recombination and desorption are suppressed) they may accumulate on the
surface and reach a high coverage.

\section{Master Equation for H$_2$ Formation on Small Grains}
\label{sec:MasterModel}

We will now consider the formation of H$_2$ molecules on small
dust grains in interstellar clouds.
In this case it is more convenient to rescale our parameters
such that instead of using quantities per unit area - the total
amount per grain will be used.
The number of H atoms on the grain is 
denoted by
$N_{\rm H}$.
Its expectation value is
given by
$\langle N_{\rm H} \rangle = S \cdot n_{\rm H}$
where $S$ is the number of adsorption sites on the grain.
Similarly, the number of H$_2$ molecules
on the grain is  
$N_{\rm H_2}$ 
and its expectation value
is  
$\langle N_{\rm H_2} \rangle = S \cdot n_{\rm H_2}$
(we assume that each adsorption site can adsorb either an H
atoms or an H$_2$ molecule).
The incoming flux of H atoms onto the grain surface is given by
$F_{\rm H} = S \cdot f_{\rm H}$ (atoms s$^{-1}$).
The desorption rates 
$W_{\rm H}$
and
$W_{\rm H_2}$
remain unchanged.
The hopping rate 
$a_{\rm H}$
(hops s$^{-1}$)
is replaced by 
$A_{\rm H} = a_{\rm H}/S$
which is approximately the inverse of the time
$t_s$
required for an atom
to visit nearly all the
adsorption sites on the grain surface.
This is due to the fact that in two dimensions the 
number of distinct sites visited by a random walker
is linearly proportional to the number of steps, up
to a logarithmic correction
\citep{Montroll1965}.
The H$_2$
production rate of the single grain
is given by
$R_{\rm H_2} = S \cdot r_{\rm H_2}$
(molecules s$^{-1}$).
The rate equations 
(neglecting the Langmuir-Hinshelwood rejection term)
will thus take the form
\begin{subeqnarray}
\label{eq:Ngrain}
{ {d{ \langle N_{\rm H} \rangle }} \over {dt}} & = & F_{\rm H} 
- W_{\rm H} \langle N_{\rm H} \rangle -
 2 A_{\rm H} {\langle N_{\rm H} \rangle}^{2} 
\slabel{eq:N1grain} \\
{ {d{\langle N_{\rm H_2} \rangle }} \over {dt}}  & = & F_{\rm H_2} 
+ \mu A_{\rm H} {\langle N_{\rm H} \rangle}^{2}
 - W_{\rm H_2} \langle N_{\rm H_2} \rangle, 
\slabel{eq:N2grain}
\end{subeqnarray}
where the first term in
(\ref{eq:N2grain})
accounts for the flux of hydrogen molecules from the gas phase
that are adsorbed on the grain surface.
While for large grains 
Eqs.~(\ref{eq:Ngrain})
provide a good description of the recombination processes,
in the limit in which the number of atoms on the grain
becomes small they are not suitable anymore. 

In order to to resolve this problem
we will now introduce a different  approach
based on a master equation that is
suitable for the study of H$_2$ formation on small grains.
Each grain is exposed to a flux $F_{\rm H}$ of H atoms.
At any given time the number of H atoms adsorbed on the grain
may be $N_{\rm H}=0, 1, 2, \dots$.
The probability that there are $N_{\rm H}$ hydrogen atoms  
on the grain 
is given by
$P_{\rm H}(N_{\rm H})$,
where
\begin{equation}
\sum_{N_{\rm H}=0}^{\infty} P_{\rm H}(N_{\rm H}) =1.
\label{eq:normalization}
\end{equation}
The time 
derivatives
of these probabilities,
$\dot P_{\rm H}(N_{\rm H})$,
are given by 
\begin{eqnarray}
\label{eq:Nmicro}
\dot P_{\rm H}(0) &=& - F_{\rm H} P_{\rm H}(0) + W_{\rm H} P_{\rm H}(1) 
+ 2 \cdot 1 \cdot A_{\rm H} P_{\rm H}(2) \nonumber \\
\dot P_{\rm H}(1) &=&  F_{\rm H} \left[ P_{\rm H}(0) - P_{\rm H}(1) \right] 
+ W_{\rm H} \left[ 2 P_{\rm H}(2) - P_{\rm H}(1) \right] 
                + 3 \cdot 2 \cdot A_{\rm H} P_{\rm H}(3) \nonumber \\
\dot P_{\rm H}(2) &=&  F_{\rm H} \left[ P_{\rm H}(1) - P_{\rm H}(2) \right] 
+ W_{\rm H} \left[ 3 P_{\rm H}(3) - 2 P_{\rm H}(2) \right] \nonumber \\
                &+& A_{\rm H} \left[ 4 \cdot 3 \cdot P_{\rm H}(4) 
                 - 2 \cdot 1 \cdot P_{\rm H}(2) \right] \nonumber \\
&\vdots& \nonumber \\
\dot P_{\rm H}(N_{\rm H}) &=&  
F_{\rm H} \left[ P_{\rm H}(N_{\rm H}-1) - P_{\rm H}(N_{\rm H}) \right] 
+ W_{\rm H} \left[ (N_{\rm H}+1) P_{\rm H}(N_{\rm H}+1)
- N_{\rm H} P_{\rm H}(N_{\rm H}) \right] \nonumber \\
                &+& A_{\rm H} \left[ (N_{\rm H}+2)(N_{\rm H}+1)
 P_{\rm H}(N_{\rm H}+2) 
-  N_{\rm H}(N_{\rm H}-1) P_{\rm H}(N_{\rm H}) \right].   \\
&\vdots& \nonumber 
\end{eqnarray}
Each of these equations includes three terms.
The first term describes the effect of the incoming flux $F_H$ on the
probabilities.
The probability $P_{\rm H}(N_{\rm H})$ increases when an H atom is
adsorbed on a grain that already
has $N_{\rm H}-1$ adsorbed atoms 
[at a rate of $F_{\rm H} P_{\rm H}(N_{\rm H}-1)$], 
and decreases when a new atom is adsorbed on a grain with
$N_{\rm H}$ atoms on it
[at a rate of $F_{\rm H} P_{\rm H}(N_{\rm H})$].
The second term includes the effect of desorption. 
An H atom desorbed from a grain with $N_{\rm H}$ adsorbed atoms decreases the
probability $P_{\rm H}(N_{\rm H})$
[at a rate of
$N_{\rm H} W_{\rm H} P_{\rm H}(N_{\rm H})$,  where the factor $N_{\rm H}$
is due to the fact that each of the
$N_{\rm H}$ atoms can desorb] 
and increases the probability
$P_{\rm H}(N_{\rm H}-1)$
at the same rate.
The third term describes the effect of recombination on the number of adsrobed
H atoms. 
The production of one molecule reduces this  number from $N_{\rm H}$ to
$N_{\rm H}-2$.
For one pair of H atoms the recombination rate is 
proportional to 
the sweeping rate
$A_{\rm H}$ 
multiplied by 2 since both atoms are mobile
simultaneously.
This rate is multiplied by
the number of possible pairs of atoms, namely
$N_{\rm H}(N_{\rm H}-1)/2$. 
Note that the equations for 
$\dot P_{\rm H}(0)$ 
and
$\dot P_{\rm H}(1)$ 
do not include all the terms, because at least one H 
atom is required for desorption to occur and at least two
for recombination.
The rate of formation of H$_2$ molecules,
$R_{\rm H_2}$ (molecules s$^{-1}$), 
is thus
given by 
\begin{equation} 
R_{\rm H_2} = A_{\rm H} \sum_{N_{\rm H}=2}^{\infty} N_{\rm H}(N_{\rm H}-1) P_{\rm H}(N_{\rm H}).
\label{eq:Rgrain}
\end{equation}
Note that the sum could start from $N_{\rm H}=0$
since the first two terms vanish.
The recombination efficiency is given by
\begin{equation}
\eta = {R_{\rm H_2} \over (F_{\rm H}/2)}.
\end{equation}
The probability that there are $N_{\rm H_2}$ hydrogen molecules 
on the grain 
is given by
$P_{\rm H_2}(N_{\rm H_2})$.
The time evolution of these probabilities
is given by
\begin{eqnarray}
\label{eq:Nmicromol}
\dot P_{\rm H_2}(0) &=& - F_{\rm H_2} P_{\rm H_2}(0) +
 W_{\rm H_2} P_{\rm H_2}(1) 
- \mu R_{\rm H_2}  P_{\rm H_2}(0) \nonumber  \\
\dot P_{\rm H_2}(1) &=&  F_{\rm H_2} \left[ P_{\rm H_2}(0)
 - P_{\rm H_2}(1) \right] 
+ W_{\rm H_2} \left[ 2 P_{\rm H_2}(2) - P_{\rm H_2}(1) \right] 
                +\mu R_{\rm H_2} \left[ P_{\rm H_2}(0) -
 P_{\rm H_2}(1) \right] \nonumber \\
\dot P_{\rm H_2}(2) &=&  F_{\rm H_2} \left[ P_{\rm H_2}(1) -
 P_{\rm H_2}(2) \right] 
+ W_{\rm H_2} \left[ 3 P_{\rm H_2}(3) - 2 P_{\rm H_2}(2) \right] 
               + \mu R_{\rm H_2} \left[ P_{\rm H_2}(1) -
 P_{\rm H_2}(2) \right] \nonumber \\
&\vdots& \nonumber \\
\dot P_{\rm H_2}(N_{\rm H_2}) &=&  F_{\rm H_2} 
\left[ P_{\rm H_2}(N_{\rm H_2}-1) - P_{\rm H_2}(N_{\rm H_2}) \right] 
+ W_{\rm H_2} \left[ (N_{\rm H_2}+1) P_{\rm H_2}(N_{\rm H_2}+1)
 - N_{\rm H_2} P_{\rm H_2}(N_{\rm H_2}) \right] \nonumber \\ 
                &+& \mu R_{\rm H_2} 
\left[ P_{\rm H_2}(N_{\rm H_2}-1) - P_{\rm H_2}(N_{\rm H_2}) \right]  \\
&\vdots& \nonumber 
\end{eqnarray}
where $F_{\rm H_2}$ 
(molecules s$^{-1}$)
is the flux of H$_2$ molecules that stick on the grain surface,
$W_{\rm H_2}$ is the desorption rate of molecules from the surface
(which is inversely proportional to their residence time, namely
$t_{\rm H_2} = 1/W_{\rm H_2})$.
Each of these equations includes three terms, describing the effects
of an incoming H$_2$ flux, desorption and recombination, respectively.

Note that the fact that we ignored the Langmuir-Hinshelwood rejection term
allowed us to split the master equation into two parts:
Eq.~(\ref{eq:Nmicro})
for the H atoms
and 
Eq.~(\ref{eq:Nmicromol})
for the H$_2$ molecules.
Moreover, 
Eq.~(\ref{eq:Nmicro})
does not depend on the distribution of 
$N_{\rm H_2}$,
while 
Eq.~(\ref{eq:Nmicromol})
depends only on the first and second moments of the
distribution of
$N_{\rm H}$.
The most general case, in which the rejection term is included,
would require to use a
master equation for the joint probability distribution
$P_{{\rm H} \& {\rm H_2}}(N_{\rm H},N_{\rm H_2})$,
which is clearly much more complicated.

The expectation value for the number of H atoms on the grain 
is  
\begin{equation}
\langle N_{\rm H} \rangle = \sum_{N_{\rm H}=0}^{\infty}
 N_{\rm H} P_{\rm H}(N_{\rm H})
\end{equation}
and the expectation value
for the number of molecules is 
\begin{equation}
\langle N_{\rm H_2} \rangle= \sum_{N_{\rm H_2}=0}^{\infty}
 N_{\rm H_2} P_{\rm H_2}(N_{\rm H_2}).
\end{equation}
The time dependence of these expectation values,
obtained from Eqs. 
(\ref{eq:Nmicro})
and 
(\ref{eq:Nmicromol}),
is given by 
\begin{eqnarray}
{d \langle N_{\rm H} \rangle \over {dt} } &=& F_{\rm H} -
 W_{\rm H} \langle N_{\rm H} \rangle 
- 2 A_{\rm H} \langle N_{\rm H}(N_{\rm H}-1) \rangle \\
{d \langle N_{\rm H_2} \rangle \over {dt} } &=& F_{\rm H_2} + 
\mu A_{\rm H} \langle N_{\rm H}(N_{\rm H}-1) \rangle
 - W_{\rm H_2} \langle N_{\rm H_2} \rangle,
\end{eqnarray}
and the recombination rate $R_{\rm H_2}$
(molecules s$^{-1}$)
for the grain is
\begin{equation}
R_{\rm H_2} = (1-\mu) A_{\rm H} \langle N_{\rm H}(N_{\rm H}-1) \rangle 
+ W_{\rm H_2} \langle N_{\rm H_2} \rangle.
\end{equation}
These equations resemble the rate equations
(\ref{eq:Ngrain})
apart from one important difference:
the recombination term
${\langle N_{\rm H} \rangle}^2$
is replaced by 
$\langle {N_{\rm H}}^2 \rangle - \langle N_{\rm H} \rangle $.
On a macroscopically large grain it is expected that the difference
between these two terms
will be small and 
Eqs.
(\ref{eq:Ngrain})
would provide a good description of the system.
However, on a small grain, where 
${\langle N_{\rm H} \rangle}$ is small these two terms are significantly
different
and it is necessary to use the 
master equation rather than the rate equations.

In principle the master equation consists of
infinitely many equations for each atomic or molecular specie.
In practice, for each specie such as atomic hydrogen
we simulate a finite number of equations for
$P_{\rm H}(N_{\rm H})$, $N_{\rm H}=1,\dots,N_{\rm max}$,
where
$P_{\rm H}(N_{\rm H})=0$ for $N_{\rm H} > N_{\rm max}$.
Obviously, $N_{\rm max}$ cannot exceed the number of adsorption sites, $S$,
on the grain.
In the equations
for
$P_H(N_{\rm max}-1)$
and for
$P_H(N_{\rm max})$,
the terms that couple them to
$P_H(N_{\rm max}+1)$
and
$P_H(N_{\rm max}+2)$
are removed
(these terms describe the flow of probability from 
$P_{\rm H}(N_{\rm H})$, $N_{\rm H} > N_{\rm max}$
to
$P_{\rm H}(N_{\rm H})$, $N_{\rm H} \le N_{\rm max}$).
Terms such as
$F_{\rm H} P_{\rm H}(N_{\rm H})$
that describe probability flow in the opposite
direction are also removed.
The latter terms are evaluated separately
and frequently during the integration of the master
equation
in order to examine whether $N_{\rm max}$
should be increased.
The condition for adding more equations is typically
$\dot P_{\rm H}(N_{\rm max}+1) > \epsilon$
at a certain time $t$,
where $\epsilon$
is a small parameter,
suitably chosen according to the desired precision.

Note that the master equation is typically
needed when 
$\langle N_{\rm H}\rangle$
is of order unity.
Under such conditions most of the probability
$P_{\rm H}(N_{\rm H})$
is concentrated at small values of $N_{\rm H}$
and therefore
$N_{\rm max}$
is expected to be small.
In a time dependent simulation,
when
$\langle N_{\rm H}\rangle$
increases reaching the limit
$\langle N_{\rm H}\rangle \gg 1$
(thus requiring $N_{\rm max} \gg 1$)
the master equation can be easily
replaced by the rate equations, during the run.
One simply has to evaluate
$\langle N_{\rm H}\rangle$ 
at a certain time $t$
and from that point to continue the run
using the rate equations.
The opposite move of switching from the rate equations
to the master equation
(when $\langle N_{\rm H}\rangle$ decreases towards $\langle N_{\rm
H}\rangle \approx 1$)
is nearly as simple.
One has to pick as an initial condition for the master equation
a narrow distribution 
$P_{\rm H}(N_{\rm H})$
that satisfies the average
$\langle N_{\rm H}\rangle$ 
given by the rate equations,
and after some relaxation time it will converge to the proper distribution.
In simulations of more complex reactions involving multiple species,
the coupling between different species typically involves only averages such as
$\langle N_{\rm H}\rangle$ (this is an approximation that will be
discussed below). 
Therefore, one can simultaneously use the rate equations for some species
and the
master equation for others, according to the criteria mentioned above.

To couple 
the master equation,
consisting of
Eqs.~(\ref{eq:Nmicro})
and 
(\ref{eq:Nmicromol}),
to the densities in the gas phase we will
consider the densities 
$\rho_H$ (atoms $cm^{-3}$) of H atoms
and 
$\rho_{\rm H_2}$ (molecules cm$^{-3}$) of H$_2$
molecules in the gas phase.
The incoming fluxes onto the surface of a single grain can be expressed as
$F_{\rm H} = \rho_H v_H \sigma$ 
and
$F_{\rm H_2} = \rho_{\rm H_2} v_{\rm H_2} \sigma$ 
where $v_{\rm H_2}$ is the average speed of an H$_2$ molecule in the gas phase.
The time derivatives of the densities are given by
\begin{eqnarray}
\dot \rho_{\rm H} &=& \left[ -F_{\rm H} + W_{\rm H}^{} \langle N_{\rm H}
\rangle \right] \rho_{\rm g} \\
\dot \rho_{\rm H_2} &=& \left[ -F_{\rm H_2}^{} + (1-\mu) A_{\rm H}
\langle N_{\rm H}(N_{\rm H}-1) \rangle 
+ W_{\rm H_2} \langle N_{\rm H_2} \rangle \right] \rho_{\rm g}.
\end{eqnarray}

\section{Simulations and Results}
\label{sec:Simulations} 

To examine the effect of the finite grain size on the recombination rate
of hydrogen in the interstellar medium we performed simulations of
the recombination process using the master equation [that consists of
Eqs.~(\ref{eq:Nmicro})
and 
(\ref{eq:Nmicromol})]
and compared the results to those obtained from the 
rate equations
(\ref{eq:Ngrain}).
Since we focus here on steady state conditions, only
the part of the master equation included in 
Eq.~(\ref{eq:Nmicro})
needs to be integrated, and the recombination
rate is given by
Eq.~(\ref{eq:Rgrain}).
For non-steady state conditions,
Eq.~(\ref{eq:Nmicromol})
would be essential 
in order to evaluate the time dependent recombination rate.
The parameters we have used are given below.
Assuming, 
for simplicity,
a spherical grain 
of diameter $d$
we obtain a cross section of $\sigma = \pi d^2/4$.
The estimate of
the number of adsorption sites on the grain  
was based on the experimental data for
the olivine and amorphous carbon surfaces
(Pirronello et al. 1997a, 1997b, 1999)
using the following procedure.
The flux of the H and D beams was estimated as
$b \cong 10^{12}$ 
(atoms cm$^{-2}$ s$^{-1}$).
The beams passed through a chopper that reduced their flux by a factor
of $c=20$.
A measurement of the flux in units of ML per second was done using the 
data for the total HD yield vs. exposure time 
[Fig. 3 in Pirronello et al. (1997a)].
The theoretical Langmuir-Hinshelwood mechanism provides a prediction
for the coverage of adsorbed atoms after 
irradiation time $t$, which is
\begin{equation}
n_{\rm H}(t) = 1 - \exp(-f_{\rm H} \cdot t).
\end{equation}
Fitting the total HD yield to this expression we obtained 
good fits that provide the
flux values 
$f_{\rm H} = 2.7 \cdot 10^{-4}$ (in ML s$^{-1}$) 
for the 
olivine experiment and 
$f_{\rm H} = 9.87 \cdot 10^{-4}$ 
for the amorphous carbon experiment.
From these two measurements we obtain the 
density of adsorption sites (sites cm$^{-2}$) 
\begin{equation}
s = {b \over {c \cdot f_{\rm H}}}.
\end{equation}
For the olivine surface it is found that 
$s \cong 2 \cdot 10^{14}$ 
and for the amorphous carbon surface
$s \cong 5 \cdot 10^{13}$
(sites cm$^{-2}$). 

Observations indicate that there is a broad distribution of grain sizes,
that roughly resembles power-law behavior,
in the range of 
$10^{-6} {\rm cm} < d < 10^{-4} {\rm cm}$
\citep{Mathis1977,Mathis1996,O'Donnell1997}.
The number of adsorption sites on a (spherical) grain is given by
$S = \pi d^2 s$.
In the simulations we focus on diffuse clouds and
use as a typical value for the 
density of H atoms 
$\rho_{\rm H} = 10$ (atoms cm$^{-3}$). 
The temperature of the H gas is taken as $T= 100$K.
The typical velocity of H atoms in the gas phase is
given by
\citep{Landau1980}
\begin{equation}
v_{\rm H} = \sqrt{  {8 \over \pi} \cdot  { {k_B T} \over {m} }  }
\end{equation}
where $m=1.67 \cdot 10^{-24}$ (gram)
is the mass of an H atom.
We thus obtain
$v_H = 1.45 \cdot 10^5$ (cm s$^{-1}$).
The density of grains is typically taken as
$\rho_{\rm g} = 10^{-12} \rho_{\rm H}$
and hence in our case
$\rho_{\rm g} = 10^{-11}$
(grains cm$^{-3}$).
The sticking probability of H atoms onto the grain surface
is taken as $\xi = 1$. 
Experimental results indicate that the sticking probability is 
close to 1 for temperatures below about $10$K and
possibly somewhat lower at higher temperatures. 
Since there is no high quality experimental data for 
the temperature dependence
$\xi(T)$, and in order to simplify the analysis
we chose
$\xi=1$.

We will now analyze the processes that take place on a single grain
using numerical integration of the master equation and comparison to
the rate equations. 
The flux of H atoms onto the grain surface is given by
$f_{\rm H} = \rho_{\rm H} v_{\rm H} / (4 s)$ (ML s$^{-1}$),
where the factor of 4 in the denominator is the ratio between
the surface area and the cross section for a spherical grain.
Using the parameters above we obtain that
$f_{\rm H} = 0.18 \cdot 10^{-8}$
for olivine and
$f_{\rm H} = 0.73 \cdot 10^{-8}$
for amorphous carbon.
The total flux on a grain of diameter $d$
is given by
$F_{\rm H} = f_{\rm H} \cdot S$ (atoms s$^{-1}$).

In Fig
\ref{fig:olivine.diffuse}
we present the recombination efficiency 
$\eta$
for
an olivine grain,
under steady state conditions, 
as a function
of the grain temperature.
The solid line shows the results obtained from the
rate equations,
showing a range of very high efficiency between
$7 - 9$ K and a tail of decreasing efficiency
between $9 - 10$ K.
The results of the master equation are shown for
grains of diameters
$d = 10^{-5} {\rm cm}$ ($\bigcirc$) 
and
$d = 10^{-6} {\rm cm}$ ($\times$).
In this case the total flux on a grain
amounts to 
$F_{\rm H} = 10^{-3}$ and $10^{-5}$ (atoms s$^{-1}$)
for the larger and smaller grain sizes, respectively. 
It is found that for the larger grain there 
is good agreement between the master equation and rate equation results.
However, for the smaller grain the master
 equation predicts significantly lower
recombination efficiency for temperatures higher than 8 K.
Note that in order to produce the rate equation results for the
entire temperature range of 5 - 15 K we had to include 
in these equations the Langmuir-Hinshelwood rejection term
\citep{Katz1999}.
In the low temperature limit, where atoms are immobile
(and the recombination rate decreases to zero), this is
necessary in order to make sure that the coverage does not exceed 1 ML.
For higher temperatures, where the comparison with the master equation
results is made, the coverage is low and the rejection term makes no
difference. 
From the results for the recombination efficiency one obtains the
production rate of H$_2$ molecules 
${\cal R}_{H_2} = {1 \over 2} F_{\rm H} \rho_{\rm g} \eta$
(molecules cm$^{-3}$ s$^{-1}$)
released to the gas phase 
by a density 
$\rho_{\rm g}$
of grains.

In Fig.
\ref{fig:olivineN_H}
we present the 
expectation value for the number 
$\langle N_H \rangle$ 
of H atoms on an olivine grain 
as a function of the grain
diameter $d$.
The results 
($\bigcirc$)
are obtained from the master equation
under steady state conditions at $T=10$K.
The solid line is simply a guide to the eye.
As may be expected,
$\langle N_H \rangle \sim d^2$, 
namely, it is proportional to the surface area
of the grain.
It is observed that for a grain diameter
smaller than about
$10^{-5}$ (cm) 
the expectation value
$\langle N_H \rangle$ 
decreases below one H atom on the grain.
Under these conditions significant deviations
are expected between the recombination efficiency predicted by the rate
equations and the correct results obtained from the master equation.

The recombination efficiency $\eta$ for
an olivine grain 
as a function
of the grain size ($\bullet$)
is shown in 
Fig.~\ref{fig:olivine.grainsize}.
The temperature and the flux are identical to 
those used in 
Fig.~\ref{fig:olivineN_H}. 
The dashed line shows the recombination efficiency obtained from
the rate equations under similar conditions, 
which is independent of the grain size.
It is observed that for 
grain diameter
smaller than about
$10^{-5}$ (cm) 
the recombination efficiency sharply drops below the rate-equation
value.
This is due to the fact that in this range
$\langle N_H \rangle < 2$,
hence most often an H atom 
resides alone
on the grain
and no recombination is possible. 

In 
Fig.~\ref{fig:olivine.N_Hdist}
we present
the distribution 
$P_{\rm H}(N_{\rm H})$
on olivine grains 
of diameters 
$d=10^{-5}$ ${\rm cm}$ ($\bigcirc$) 
and
$d=10^{-6}$ ${\rm cm}$ ($\times$). 
The results were
obtained from the master equation
under steady state conditions 
at $T=9$ K.
For the larger grain
the distribution exhibits a peak around 
$\langle N_H \rangle \cong 14$
and the rate equations are expected to apply.
However, for the smaller grain
the highest probability is for having no H atoms
at all on the grain and
$\langle N_H \rangle < 1$.
Under these conditions 
the rate equations are expected to highly over-estimate
the recombination efficiency.
Indeed, this can be observed in 
Fig.~\ref{fig:olivine.diffuse},
in which the recombination efficiency for
the smaller grain size at 9 K is much lower
than the rate equation result.

The results for the recombination efficiency on amorphous
carbon grains are qualitatively similar to those shown here
for olivine. 
The temperature range of very
high efficiency is between 
$12 - 16$ K and the tail of decreasing efficiency
is between 
$16 - 18$ K
[see Figs.~6 and 7 in Katz et al. (1999)].
This high efficiency window is within the relevant
temperature range for diffuse cloud environments.
It is observed that
for temperatures higher than about 15 K and grain sizes
smaller than
$10^{-5}$ (cm)
the master equation predicts lower recombination efficiency 
than the rate equations.

\section{More Complex Reactions of Multiple Species}
\label{sec:Complex}

The master equation introduced above can be extended to describe
more complex situations that involve chemical reactions with multiple
species. 
We chose to demonstrate this procedure for the reactions involving
oxygen and hydrogen on dust grains, 
previously studied by
Caselli et al. (1998).
The rate equations that describe these reactions are
\begin{subeqnarray}
\label{eq:NgrainHO}
{ {d{\langle N_{\rm H} \rangle}} \over {dt}} & = & F_{\rm H} - W_{\rm H}
\langle N_{\rm H} \rangle 
- 2 A_{\rm H} {\langle N_{\rm H} \rangle}^{2}  \nonumber \\
                             & - & (A_{\rm H}+A_{\rm O}) \langle N_{\rm
H}\rangle \langle N_{\rm O} \rangle 
- A_{\rm H} \langle N_{\rm H} \rangle \langle N_{\rm OH} \rangle   
\slabel{eq:N1grainHO} \\
{ {d{\langle N_{\rm O} \rangle }} \over {dt}} & = & F_{\rm O} -  (A_{\rm
H}+A_{\rm O}) \langle N_{\rm H} \rangle \langle N_{\rm O} \rangle - 2
A_{\rm O} {\langle N_{\rm O} \rangle}^{2} 
\slabel{eq:N2grainHO} \\
{ {d{\langle N_{\rm H_2} \rangle }} \over {dt}}  & = & \mu A_{\rm H}
{\langle N_{\rm H} \rangle}^{2} 
- W_{\rm H_2} \langle N_{\rm H_2} \rangle. 
\slabel{eq:N3grainHO} \\ 
{ {d{\langle N_{\rm OH} \rangle }} \over {dt}}  & = &  
(A_{\rm H}+A_{\rm O}) \langle N_{\rm H} \rangle \langle N_{\rm O} \rangle
- A_{\rm H} \langle N_{\rm H} \rangle \langle N_{\rm OH} \rangle  
\slabel{eq:N4grainHO} \\
{ {d{\langle N_{\rm O_2} \rangle }} \over {dt}}  & = &  A_{\rm O}
{\langle N_{\rm O} \rangle}^2 
\slabel{eq:N5grainHO} \\
{ {d{\langle N_{\rm H_2O} \rangle}} \over {dt}}  & = &  A_{\rm H} \langle
N_{\rm OH} \rangle \langle N_{\rm H} \rangle 
\slabel{eq:N6grainHO}
\end{subeqnarray}
where
$N_{\rm O}$ is the number of oxygen atoms on the grain
and $A_{\rm O}$ is their sweeping rate.
The numbers of 
H$_2$, OH, O$_2$ and H$_2$O molecules on the grain are given by
$N_{\rm H_2}$, $N_{\rm OH}$, $N_{\rm O_2}$ and $N_{\rm H_2O}$,
respectively.
The flux of O atoms adsorbed on the grain surface is given by $F_{\rm O}$ 
(atoms s$^{-1}$)
while, for simplicity, the adsorption of the four molecular species from
the gas phase
is neglected.
Since the oxygen atoms and the molecules they form are chemisorbed on
 the surface they are unlikely to desorb for the surface temperatures
considered here.
Their desorption coefficients are thus neglected.
It is also assumed that the diffusion of the four molecular species is
negligible.
Except for the OH molecule, this assumption is inconsequential since
 the three other molecular species do not participate in any subsequent
reactions.
 Furthermore, it is typically assumed that the diffusion rate of oxygen
atoms is much
slower than of hydrogen, namely $A_{\rm O} \ll A_{\rm H}$. 
In the analysis below we will assume that $A_{\rm O} = 0$,
namely the reaction between H and O is driven only by the diffusion rate
of the hydrogen atoms.
Under this assumption and for low coverage of O atoms on the grain, 
the production of oxygen molecules is suppressed
and Eq.
(\ref{eq:N5grainHO})
can be ignored.
The  assumption that H atoms are much more mobile than heavier atomic
species such as
C, O  and N implies that the diffusion of H atoms is dominant in other
reactions
that take place on grain surfaces besides hydrogen recombination.
Thus, the activation energies obtained by Katz et al. (1999) 
may be used
not only for the hydrogen
 recombination process but for a large number of other reactions on dust
grain surfaces.
 However, one should carefully examine the possibility that some of these
reactions  
involve an additional activation energy associated with the reaction itself.

For  simplicity we will now neglect the reaction between H atoms and OH
molecules
which generates H$_2$O molecules. 
We will also assume, for simplicity, that H$_2$
molecules desorb into the gas phase immediately upon formation,
namely that $\mu=0$.
Under these conditions the rate equations will simplify into
\begin{subeqnarray}
\label{eq:NgrainHOs}
{ {d{\langle N_{\rm H} \rangle}} \over {dt}} & = & 
F_{\rm H} - W_{\rm H} \langle N_{\rm H} \rangle 
- 2 A_{\rm H} {\langle N_{\rm H} \rangle}^{2}  
   -  A_{\rm H} \langle N_{\rm H}\rangle \langle N_{\rm O} \rangle 
\slabel{eq:N1grainHOs} \\
{ {d{\langle N_{\rm O} \rangle }} \over {dt}} & = & 
F_{\rm O} -  A_{\rm H} \langle N_{\rm H} \rangle \langle N_{\rm O} \rangle 
\slabel{eq:N2grainHOs} \\
{ {d{\langle N_{\rm OH} \rangle }} \over {dt}}  & = &  
A_{\rm H} \langle N_{\rm H} \rangle \langle N_{\rm O} \rangle. 
\slabel{eq:N4grainHOs} 
\end{subeqnarray}
The part of the master equation describing the time evolution of the population
of H atoms on the grain surface takes the form
\begin{eqnarray}
\dot P_{\rm H}(0) &=& - F_{\rm H} P_{\rm H}(0) + W_{\rm H} P_{\rm H}(1) 
+ 2 \cdot 1 \cdot A_{\rm H} P_{\rm H}(2) 
+ A_{\rm H} P_{\rm H}(1) \langle N_{\rm O} \rangle \nonumber \\
\dot P_{\rm H}(1) &=&  F_{\rm H} \left[ P_{\rm H}(0) - P_{\rm H}(1) \right] 
+ W_{\rm H} \left[ 2 P_{\rm H}(2) - P_{\rm H}(1) \right] \nonumber \\ 
                  &+& 3 \cdot 2 \cdot A_{\rm H} P_{\rm H}(3) + 
 A_{\rm H} \left[2 P_{\rm H}(2) - P_{\rm H}(1) \right] \langle N_{\rm O}
\rangle  \nonumber \\
\dot P_{\rm H}(2) &=&  F_{\rm H} \left[ P_{\rm H}(1) - P_{\rm H}(2) \right] 
+ W_{\rm H} \left[ 3 P_{\rm H}(3) - 2 P_{\rm H}(2) \right] \nonumber \\
                 &+& A_{\rm H} \left[ 4 \cdot 3 \cdot P_{\rm H}(4) - 2
\cdot 1 \cdot P_{\rm H}(2) \right] 
 + A_{\rm H} \left[3 P_{\rm H}(3) - 2 P_{\rm H}(2) \right] \langle N_{\rm
O} \rangle \nonumber \\
&\vdots& \nonumber \\
 \dot P_{\rm H}(N_{\rm H}) &=&  F_{\rm H} \left[ P_{\rm H}(N_{\rm H}-1) -
P_{\rm H}(N_{\rm H}) \right] 
 + W_{\rm H} \left[ (N_{\rm H}+1) P_{\rm H}(N_{\rm H}+1) - N_{\rm H}
P_{\rm H}(N_{\rm H}) \right] \nonumber \\
                 &+& A_{\rm H} \left[ (N_{\rm H}+2)(N_{\rm H}+1) P_{\rm
H}(N_{\rm H}+2) 
-  N_{\rm H}(N_{\rm H}-1) P_{\rm H}(N_{\rm H}) \right] \nonumber \\
                &+& A_{\rm H} \left[(N_{\rm H}+1) P_{\rm H}(N_{\rm H}+1) 
- N_{\rm H} P_{\rm H}(N_{\rm H}) \right] \langle N_{\rm O} \rangle   \\
&\vdots& \nonumber 
\label{eq:NmicroH}
\end{eqnarray}
where
\begin{equation}
 \langle N_{\rm O} \rangle = \sum_{N_{\rm O}=1}^{\infty} N_{\rm O} P_{\rm
O}(N_{\rm O}).
\end{equation} 
The equations for oxygen atoms are
\begin{eqnarray}
\dot P_O(0) &=& - F_O P_H(0)   
+ A_{\rm H} P_O(1)  \langle N_{\rm H}\rangle \nonumber \\
\dot P_O(1) &=&  F_O \left[ P_O(0) - P_O(1) \right] 
 + A_{\rm H} \left[2 P_O(2) - P_O(1) \right] \langle N_{\rm H}\rangle
\nonumber \\
\dot P_O(2) &=&  F_O \left[ P_O(1) - P_O(2) \right] 
+ A_{\rm H} \left[3 P_O(3) - 2 P_O(2) \right] \langle N_{\rm H}\rangle    \\
&\vdots& \nonumber \\
\dot P_O(N_{\rm O}) &=&  F_O \left[ P_O(N_{\rm O}-1) - P_O(N_{\rm O}) \right]   
 + A_{\rm H} \left[(N_{\rm O}+1) P_O(N_{\rm O}+1) - N_{\rm O} P_O(N_{\rm
O}) \right] \langle N_{\rm H}\rangle . \nonumber \\
&\vdots& \nonumber 
\label{eq:NmicroO}
\end{eqnarray}
where
\begin{equation}
 \langle N_{\rm H} \rangle = \sum_{N_{\rm H}=1}^{\infty} N_{\rm H}
P_H(N_{\rm H}).
\end{equation}
The equations describing the distribution of the number of OH molecules are
\begin{eqnarray}
\dot P_{\rm OH}(0) &=& 
- A_{\rm H} P_{\rm OH}(0) 
\langle N_{\rm O}\rangle \langle N_{\rm H}\rangle  \nonumber \\
\dot P_{\rm OH}(1) &=&  
+ A_{\rm H} \left[P_{\rm OH}(0) - P_{\rm OH}(1) \right]  
\langle N_{\rm O}\rangle \langle N_{\rm H}\rangle  \nonumber \\
\dot P_{\rm OH}(2) &=&  
+ A_{\rm H} \left[P_{\rm OH}(1) - P_{\rm OH}(2) \right] 
\langle N_{\rm O}\rangle \langle N_{\rm H}\rangle  \\
&\vdots& \nonumber \\
\dot P_{\rm OH}(N_{\rm OH}) &=&  
+ A_{\rm H} \left[P_{\rm OH}(N_{\rm OH}-1) - P_{\rm OH}(N_{\rm OH}) \right] 
\langle N_{\rm O}\rangle \langle N_{\rm H}\rangle. \nonumber \\
&\vdots& \nonumber 
\label{eq:NmicroHO}
\end{eqnarray}
The rate of formation of H$_2$ molecules is given by 
Eq.~(\ref{eq:Rgrain}).
The rate of formation of OH molecules is given by
$R_{\rm OH} = A_{\rm H} \langle N_{\rm O} \rangle \langle N_{\rm H} \rangle$. 
Unlike the H$_2$ molecules that desorb, the OH 
as well as H$_2$O molecules (not included in the master equation above)
are believed to stick to the grain and form an ice mantle.
 The parameters of the bare surface, used in this paper, are suitable only
in the
early stages before the first monolayer of ice is formed.
Beyond this stage one needs the activation energies
 for H diffusion on ice, which can be obtained from TPD experiments of
hydrogen 
recombination on ice mantles.

Note that in the equations above the coupling between different
species is only through the expectation values 
$\langle N_{\rm O}\rangle$ 
and
$\langle N_{\rm H}\rangle$.
Therefore, one can simultaneously use the rate equations for some of the 
species and the master equation others according to the criteria discussed
above.
This is, in fact, an approximation in which the correlation between the 
probability distributions of different species 
such as 
$P_{\rm O}(N_{\rm O})$
and 
$P_{\rm H}(N_{\rm H})$ 
is neglected.
In cases where significant correlation is expected one can use a set of
master equation for the joint probability distribution
$P_{{\rm H} \& {\rm O}}(N_{\rm O},N_{\rm H})$.
However, in this case the number of equations that are needed is
$N_{\rm max}(O) \cdot N_{\rm max}(H)$,
which may become impractical for systems with a larger number of species.
In principle, such correlations could be significant in cases in which
there are two species that react only with each other and thus the density
of one specie would be strongly dependent on the availability of the other.
 However, in the interstellar medium such correlations are not expected to
be strong since the
 reactive species (particularly atomic hydrogen) react with a large number
of species. 

\section{Summary}
\label{sec:Summary}

 We have introduced a new approach for the simulation of hydrogen
recombination
on microscopic dust grains in the interstellar medium.
This approach is based on a set of master equation
for the probabilities
$P_{\rm H}(N_{\rm H})$, $N_{\rm H}=0,1,2 \dots$
that there are $N_{\rm H}$ hydrogen atoms on the grain surface. 
Unlike the rate equations that provide a mean-field
analysis, suitable for macroscopic surfaces,
these rate equations
provide an exact description of the recombination process
on small grains taking into account the discrete nature
of $N_{\rm H}$ as well as the fluctuations. 
The approach can be extended to more complex chemical reactions
with multiple species and coupled to the densities of the reactants
in the gas phase.

\section{Acknowledgments}
\label{sec:Acknowledgments}

We thank Eric Herbst for useful discussions that greatly stimulated
this work.
G.V. acknowledges support from NASA grant NAG5-6822.
V.P. acknowledges support from the Italian Ministry for the University and 
Scientific Research.

%\newpage

\begin{center}
Figure Captions
\end{center}

\figcaption{
The recombination efficiency $\eta$, obtained from the master equation, 
for olivine grains of diameters 
$d=0.1 \mu {\rm m}$ 
and
$d=0.01 \mu {\rm m}$ ($\times$)
under steady state conditions at a constant flux,
as a function of the grain temperature.
The flux is
$f=0.18 \times 10^{-8}$ $ML s^{-1}$,
which amounts to 
$F_{\rm H} = 10^{-3}$ and $10^{-5}$ (atoms s$^{-1}$)
for the larger and smaller grain sizes, respectively. 
The recombination efficiency, obtained from the
rate equations under similar conditions is also shown (solid line)
showing a window of high efficiency between 7 - 9 K and a tail
of decreasing efficiency above 9 K.
We observe that
in the center of the high efficiency peak there
is good agreement between the master equation and rate equation results.
However, 
for the smaller grain 
at higher temperatures,
the master equation predicts significantly lower
recombination efficiencies than the rate equations.
These deviations persist at temperatures as high as 11 and 12 K
but cannot be seen in the Figure due to the limited resolution.
\label{fig:olivine.diffuse}
}

\figcaption{
The expectation value for the number 
$\langle N_H \rangle$ 
of H atoms on an olivine grain 
as a function of the grain
diameter under
steady state conditions at $T=10$K,
as obtained from the master equation
($\bigcirc$).
The solid line is simply a guide to the eye.
\label{fig:olivineN_H}
}

\figcaption{
The recombination efficiency $\eta$ for
an olivine grain 
as a function
of the grain size ($\bullet$).
at $T=10$K
and 
$f=0.18 \times 10^{-8}$ $ML s^{-1}$.
The results predicted by the rate equations 
under similar conditions,
which are independent of the grain size,
are also shown
(dashed line).
It is observed that as the grain size decreases below
$d=0.1 \mu {\rm m}$ 
the recombination efficiency quickly decreases.
\label{fig:olivine.grainsize}
}

\figcaption{
The distributions 
$P_{\rm H}(N_{\rm H})$
on olivine grains 
of diameters 
$d=0.1$ $\mu {\rm m}$ ($\bigcirc$) 
and
$d=0.01$ $\mu {\rm m}$ ($\times$), 
obtained from the master equation
at $T=9$K
and 
$f=0.18 \times 10^{-8}$ $ML s^{-1}$. 
\label{fig:olivine.N_Hdist}
} 
    
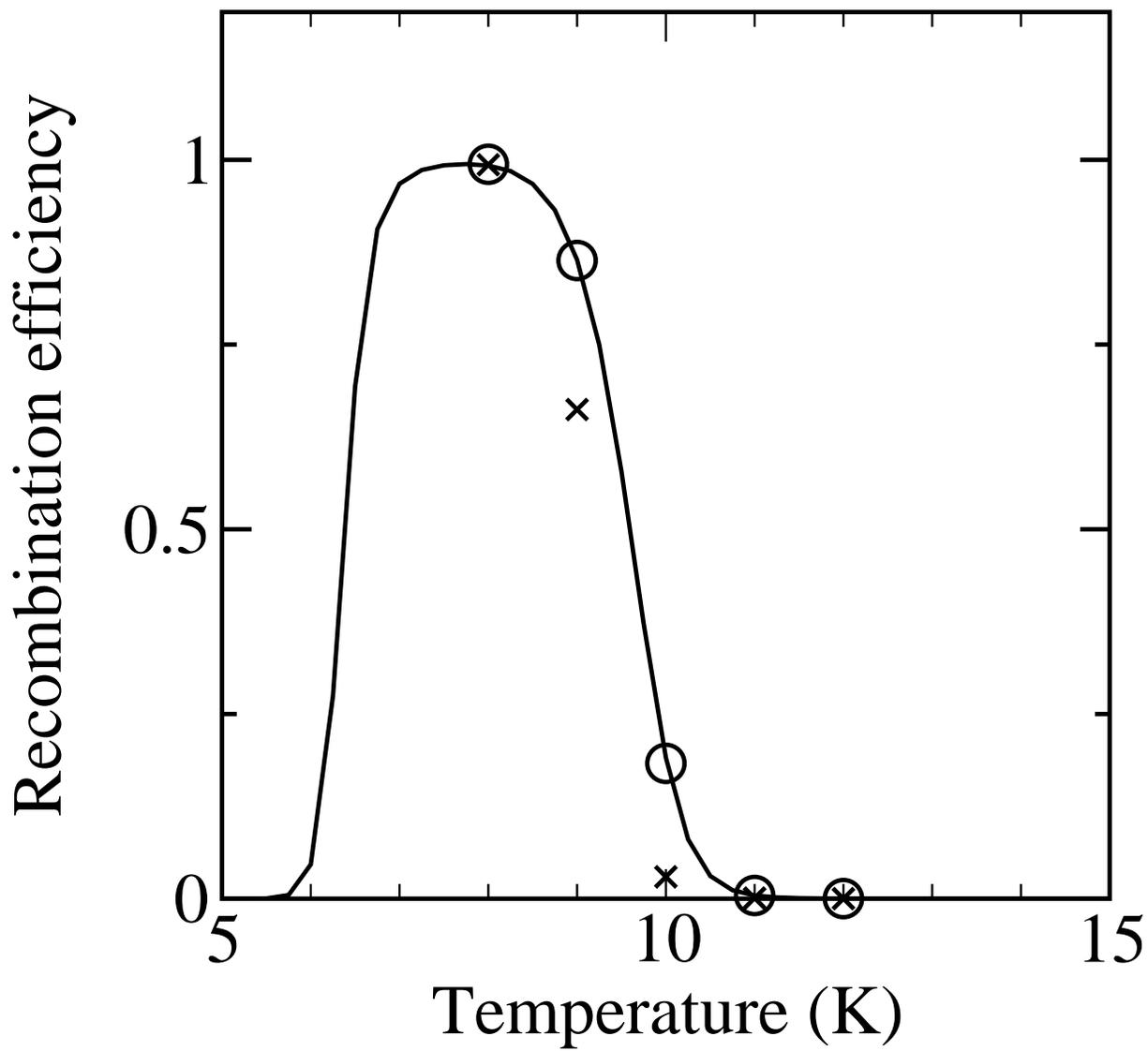
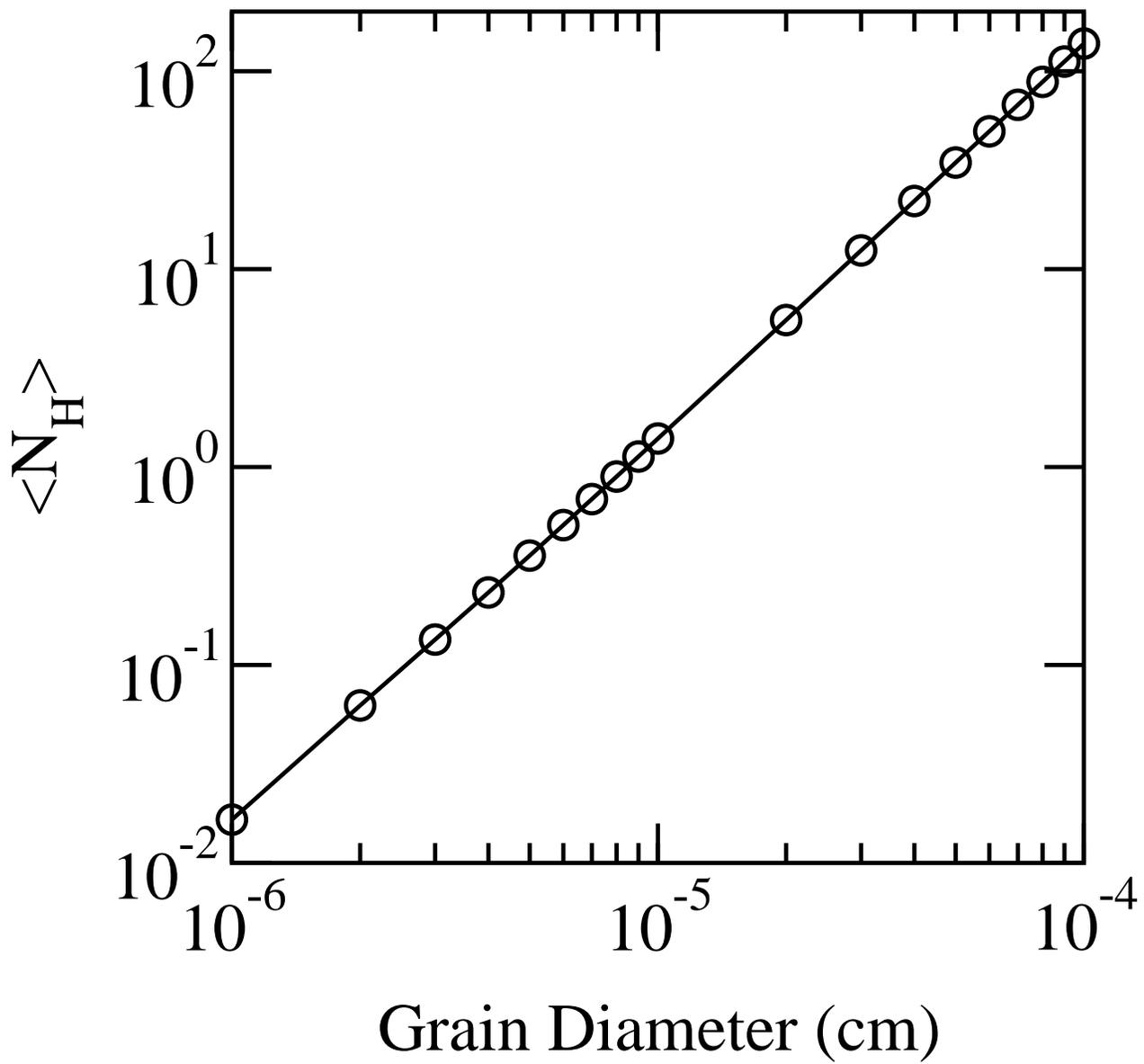
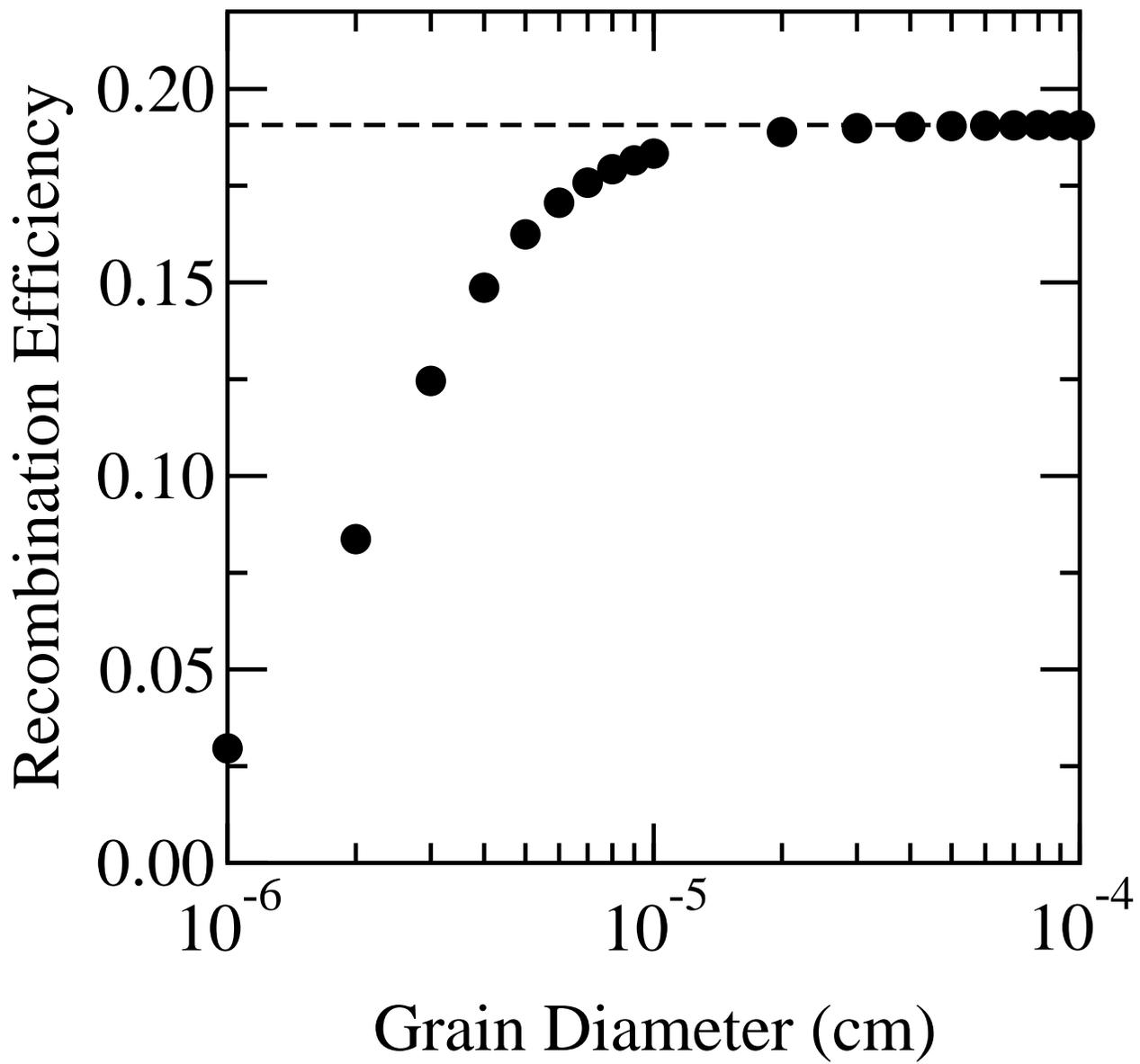
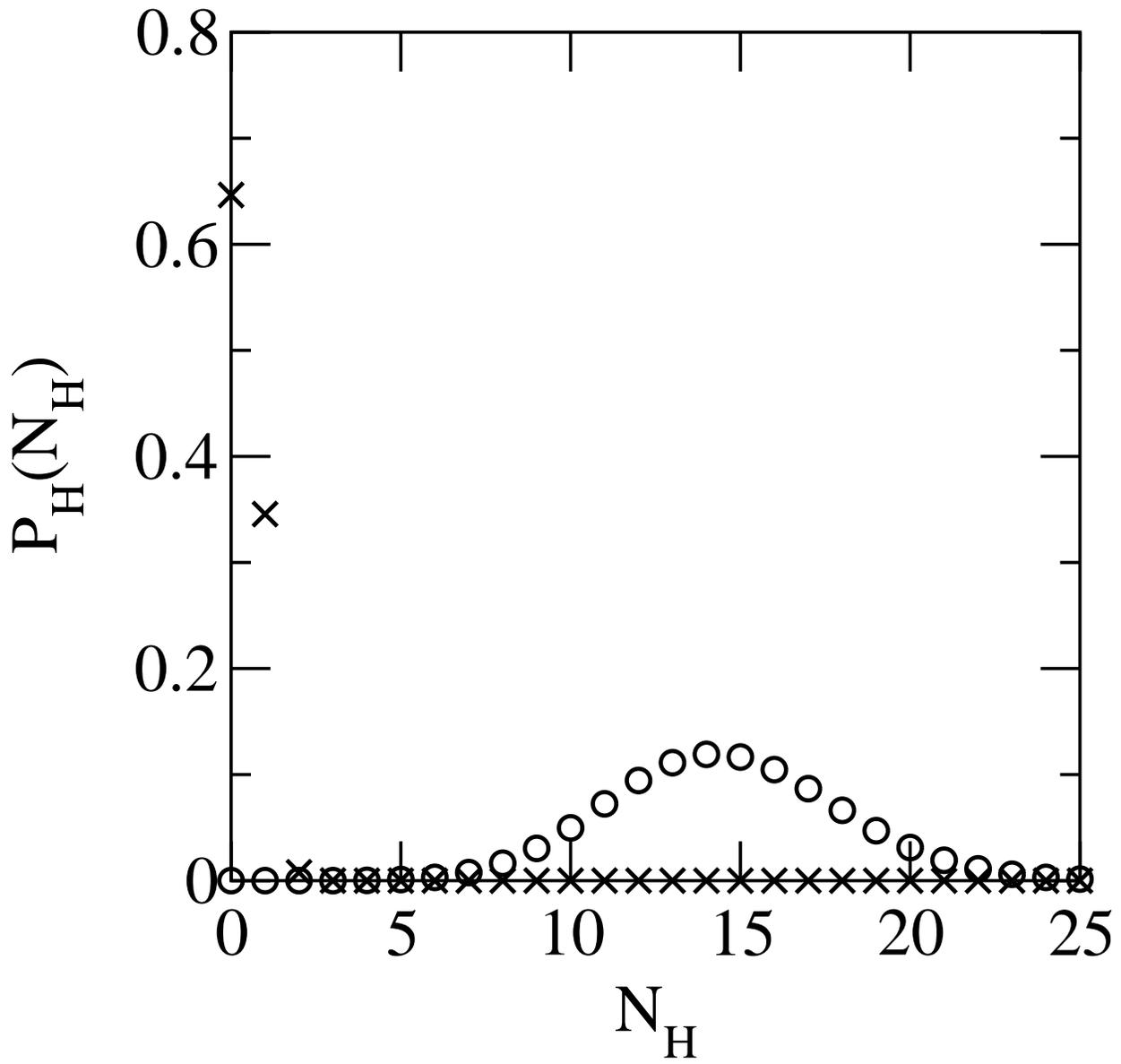
\end{document}